\documentclass[12pt]{iopart}
\usepackage{citesort}
\usepackage{braket}
\usepackage{graphicx}
\usepackage{textgreek}

\newcommand{\ketbra}[1]{{|{#1}\rangle}\!{\langle{#1}|}}

\begin{document}
\title[]{Quantum tomography of noisy ion-based qudits}

\author{B.I.~Bantysh, Yu.I.~Bogdanov}

\address{Valiev Institute of Physics and Technology of Russian Academy of Sciences, Moscow, Russia}
\ead{bbantysh60000@gmail.com}
\vspace{10pt}
\begin{indented}
\item[]November 2020
\end{indented}

\begin{abstract}
Quantum tomography makes it possible to obtain comprehensive information about certain logical elements of a quantum computer. In this regard, it is a promising tool for debugging quantum computers. The practical application of tomography, however, is still limited by systematic measurement errors. Their main source are errors in the quantum state preparation and measurement procedures. In this work, we investigate the possibility of suppressing these errors in the case of ion-based qudits. First, we will show that one can construct a quantum measurement protocol that contains no more than a single quantum operation in each measurement circuit. Such a protocol is more robust to errors than the measurements in mutually unbiased bases, where the number of operations increases in proportion to the square of the qudit dimension. After that, we will demonstrate the possibility of determining and accounting for the state initialization and readout errors. Together, the measures described can significantly improve the accuracy of quantum tomography of real ion-based qudits.
\end{abstract}
\maketitle

\section{Introduction}

The limiting factor of quantum tomography is the presence of systematic errors of various nature \cite{banaszek1998,bantysh2019,dariano2003,hou2016,bantysh2020,avosopiants2018,kimmel2014,schwemmer2015,hlousek2019,keith2018}. These errors are mainly due to the fact that the theoretical measurement model does not correspond to the real measurements. Having a model close to a real experiment, it is possible to obtain accuracy limited only by statistical fluctuations, which are lower, the higher the total sample size is \cite{bogdanov2011,gill2000}.

In conditions of statistical homogeneity of data over time, the main source of systematic errors are preparation and measurement (SPAM) errors \cite{bantysh2019,palmieri2020,magesan2012,merkel2013,huang2019,ferrie2014}. In particular, quantum state tomography (QST) implies the ability to measure an unknown state in an arbitrary basis. A set of such bases form a measurement protocol for QST \cite{bogdanov2011,bogdanov2011_2}. For the vast majority of quantum computing platforms, each measurement is performed by a basis change transformation and readout in the computational basis. Both of these procedures are carried out with errors.

Quantum process tomography (QPT) requires the ability to prepare predetermined set of states. Together with a set of measurements performed at the output of a quantum process, these states form a measurement protocol for QPT \cite{mohseni2008,bogdanov2013}. The preparation procedure is usually reduced to the initialization of the system in the $\ket0$ state and its subsequent transformation. Both initialization and transformation are error prone.

One way to suppress systematic errors in qudit tomography is to choose optimal measurement protocols, which are themselves characterized by lower error rates. For qudit tomography one usually chooses protocols with high symmetry \cite{thew2002,bent2015,lima2011,bogdanov2004,medendorp2011,varga2018}. In systems based on spatial states of light, the basis change transformation is performed by just a single operation (for example, using a spatial light modulator \cite{bent2015,varga2018}). In ion-based qudits, however, this requires a variety of 2-level transitions (\Sref{sect:ion}). In \Sref{sect:2level}, we show that for ion-based qudit state tomography, it may be sufficient to use a measurement protocol containing no more than a single elementary operation in each measurement circuit.

In practice, it is not possible to completely suppress SPAM errors. Therefore, one should use data processing methods that are robust to existing errors. Some types of errors can be accurately determined from physical considerations and taken into account in the model. For example, a standard practice is to take into account the finite quantum efficiency of photo detectors \cite{avosopiants2018,lvovsky2004,bogdanov2018,dariano2007}. However, many other error types need to be determined experimentally from the results of a set of measurements.

A promising tool here is the gate set tomography (GST), aimed at the simultaneous reconstruction of all the unknown parameters of a set of quantum gates and the SPAM system \cite{matteo2020,blume_kohout2017,rudnicki2018}. This method, however, features a fundamental ambiguity of the results. This ambiguity allows one to determine the parameters only up to a certain gauge transformation. The empty gate tomography approach \cite{bantysh2019} does not have such a drawback, but it is fundamentally limited and shows high efficiency only for SPAM errors of a certain type. It is also worth noting machine learning methods that automatically train a model on noisy data \cite{palmieri2020,fastovets2019,torlai2018}. The disadvantage here, however, is that the training stage requires the possibility of highly accurate preparation of some broad set of quantum states, which is not always possible.

The measurement protocol proposed in \Sref{sect:2level} can significantly reduce the systematic tomography errors associated with the qudit transformation errors. Thus, in \Sref{sect:init_readout}, we place emphasis on the possibility of estimating the parameters of quantum state initialization and readout. We show that a very general parametrization of these errors does not allow one to give their unambiguous estimate. In this regard, we propose to use a nonlinear model with a small number of parameters. This allowed us to obtain an unambiguous estimate and, as a result, a quantum tomography model that provided a high accuracy of reconstruction.

\section{Measurement of an ion-based qudit}\label{sect:ion}

\begin{figure}[h]
  \centering
  \includegraphics[width=\linewidth]{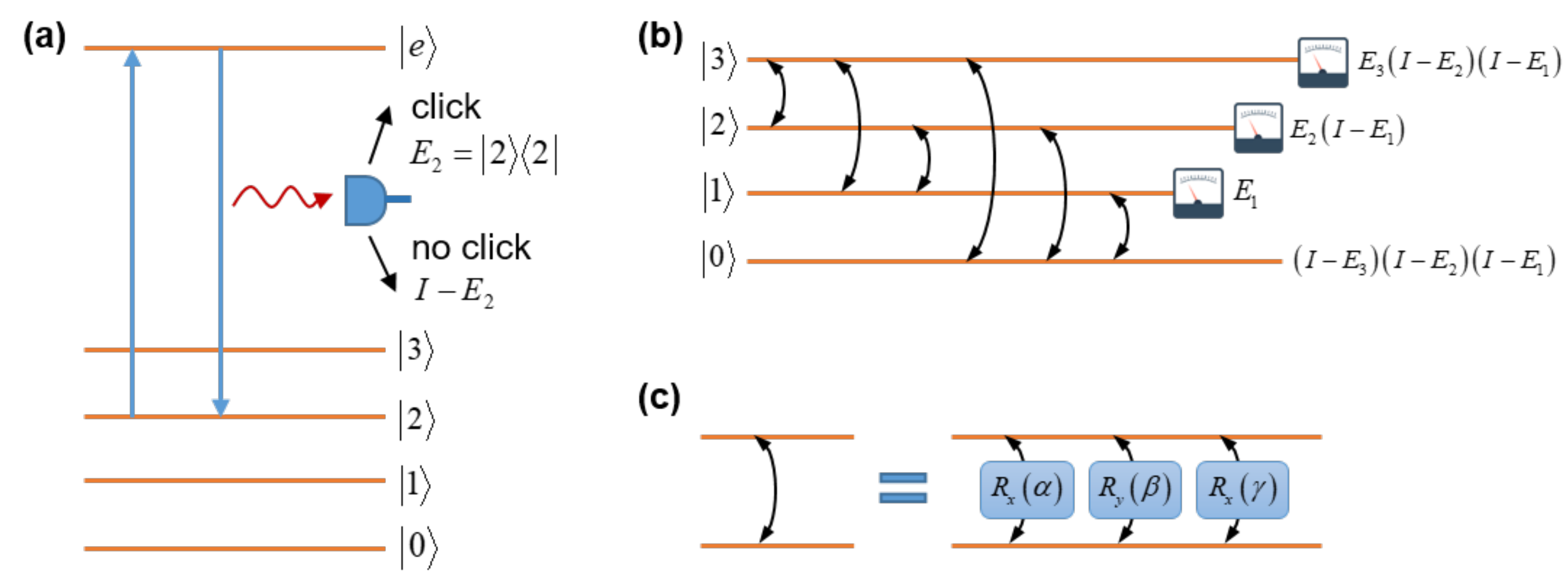}
  \caption{Measurement of an ion-based qudit ($d = 4$). (a) Level $\ket2$ population readout with the auxiliary level $\ket e$. The photon detection corresponds to the measurement operator $E_2$. (b) Qudit state measurement in an arbitrary basis. (c) An arbitrary 2-level transition is performed through the three elementary operations.}
  \label{fig:qudit_measurement}
\end{figure}

Let $d$ be the qudit dimension. The ion-based qudit measurement in the computational basis is made by sequential population readout of each individual qudit level. The $j$-th level is read by applying a resonant pulse at the frequency corresponding to the transition to the level $\ket e$ (\Fref{fig:qudit_measurement}(a)). The emitted photon could be detected by a photo detector. This event corresponds to a measurement with a projector $E_j = \ketbra{j}$. If the photon is not detected, then the event corresponds to the measurement operator $I - E_j$, where $I$ is an identity operator.

This procedure is carried out sequentially over each level from 1 to $d-1$ until the detector is triggered \cite{low2020}. The result can be described by the following set of POVM operators:
\begin{equation}\label{eq:readout_povm}
  \eqalign{\Pi_1 = E_1, \cr
  \Pi_k = E_k(I-E_{k-1})\dots(I-E_1), \quad k = 2, \dots, d-1, \cr
  \Pi_0 = (I-E_{d-1})\dots(I-E_1).}
\end{equation}
In the case of ideal measurements, such a set gives projectors ${\Pi_k = \ketbra{k}}$ (${k = 0, \dots, d - 1}$) onto the states of the computational basis.

To perform the measurement in an arbitrary basis, a unitary transformation should be performed over a qudit before readout. In general, it requires $d(d-1)/2$ 2-level transitions \cite{low2020,barenco1995} (\Fref{fig:qudit_measurement}(b)). In Euler parametrization, each such transition requires up to three elementary operations\footnote[1]{The parameterization in terms of the Euler angles actually has the form  $R_z(\gamma)R_x(\beta)R_z(\alpha)$, where $R_z$ denotes the rotation around the $z$ axis, but here we use another equivalent form for convenience}: $R_x(\gamma)R_y(\beta)R_x(\alpha)$, where $R_x$ and $R_y$ are rotation operators on a two-level Bloch sphere around the $x$ and $y$ axes, respectively (\Fref{fig:qudit_measurement}(c)).

The preparation of arbitrary states from the initial state $\ket0$ of logical zero is performed by $d-1$ 2-level transitions between $\ket0$ and other logical states.

\section{Robust qudit tomography protocol}\label{sect:2level}

It was shown above that in order to implement arbitrary QST and QPT protocols over a qudit, a large number of elementary 2-level operations must be implemented in each measurement circuit. Since each operation in the experiment is error prone, such measurement protocols can result in significant systematic errors of quantum tomography.

Hence, the problem of robust protocol construction arises. Such a protocol would require the smallest possible number of elementary operations on a qudit. It turns out that for QST it is sufficient to use no more than a single elementary operation in each measurement circuit. The first measurement basis is the computational one and does not require any operations. The following $d(d-1)/2$ bases are determined by performing operation $R_y(\pi/2)$ between each pair of qudit levels. The rest of $d(d-1)/2$ bases --- by the operation $R_x(3\pi/2)$. This choice of operations is due to the fact that in the case $d = 2$ the protocol is equivalent to the mutually unbiased bases (MUB) protocol.

We refer to this protocol as ``2-level'', since during the measurement, the amplitudes are redistributed only between two levels of a qudit. Note that such a protocol is informationally complete (the completeness criterion is from \cite{bogdanov2011}).

\begin{figure}[h]
  \centering
  \includegraphics[width=\linewidth]{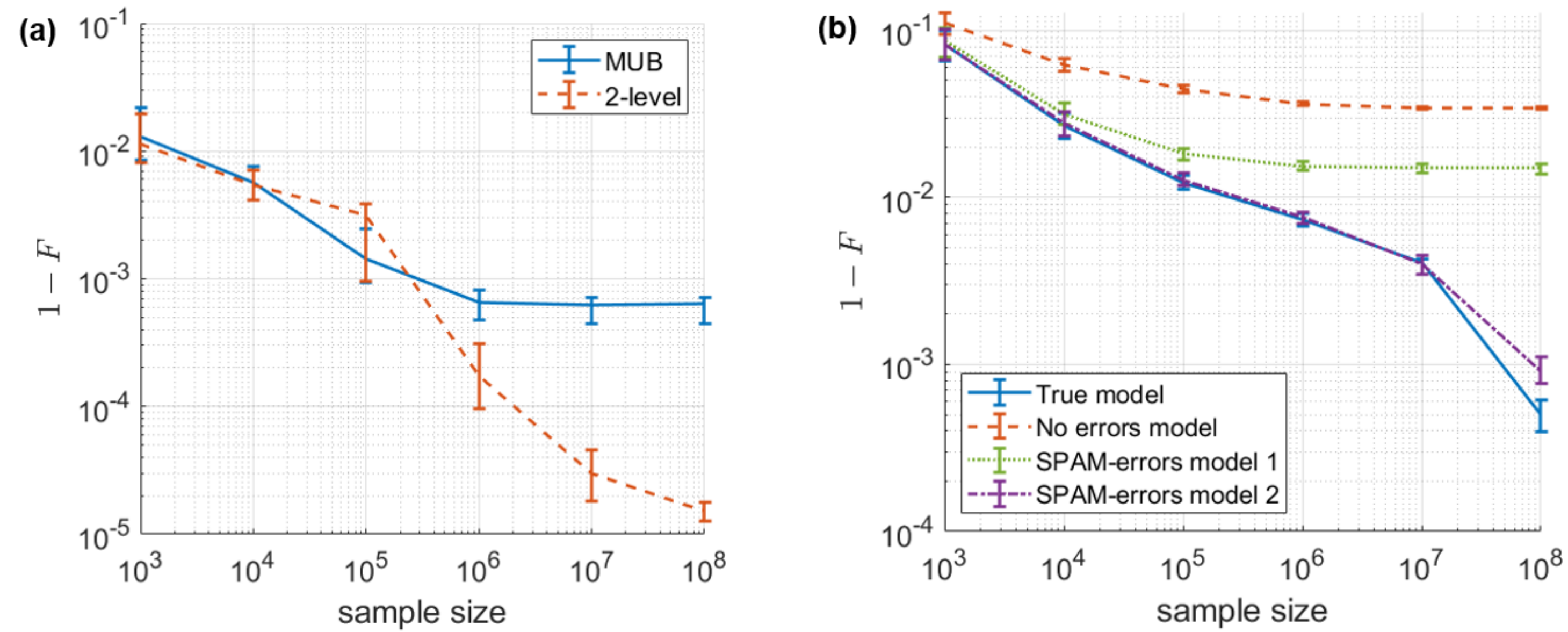}
  \caption{Average qutrit ($d=3$) tomography infidelity versus total sample size. At each point, 500 numerical experiments were performed. The statistical data were generated by the Monte Carlo method. Confidence intervals are given by the upper and lower quartiles. Each elementary 2-level unitary operation in the measurement protocol was accompanied by the channel $\mathcal{E}_{0.001}$. (a) Tomography of quantum states of the form $\mathcal{E}_{0.01}(\ketbra\psi)$, where $\ket\psi$ in each experiment were generated randomly according to the Haar measure. Two QST protocols are compared. (b) Tomography of quantum processes of the form $\mathcal{E}_{0.01}\circ\mathcal{U}$, where the unitary process $\mathcal{U}$ in each experiment was generated randomly according to the Haar measure. Various tomography models based on a 2-level protocol are compared. The graphs' kinks are due to the peculiarities of quantum state reconstruction for almost pure state \cite{bantysh2020_2}.}
  \label{fig:results}
\end{figure}

To compare protocols, consider the depolarizing quantum channel with error probability $p$:
\begin{equation}
  \mathcal{E}_p(\rho) = (1-p)\rho + pI/d.
\end{equation}
\Fref{fig:results}(a) shows the infidelity of QST of a qutrit ($d = 3$) with the states of the form $\mathcal{E}_{0.01}(\ketbra\psi)$, where $\ket\psi$ is a random (according to Haar measure) pure state. MUB and 2-level tomography protocols are considered. In the simulation, it was assumed that each ideal elementary 2-level unitary operation is accompanied by the action of depolarizing channel with the error probability $p = 0.001$. The readout is noise-free at the current stage of the simulation. For the quantum state reconstruction, we have used our open MATLAB library \cite{bantysh_root}. For small enough sample sizes, the MUB protocol yields higher fidelity than the 2-level one. This is due to the fact that this protocol carries more information about the state, and systematic errors of basis change operations are still lower than statistical fluctuations. At higher sample sizes, the 2-level protocol, being more robust to systematic errors, gives a significantly higher accuracy.

For QPT, the 2-level protocol has the following structure. The first $2(d-1)+1$ states at the process input are described by the initialization state $\ket0$ itself and $2(d-1)$ states resulting from operations $R_y(\pi/2)$ and $R_x(3\pi/2)$ between levels $\ket0$ and $\ket j$ ($j=1,\dots,d-1$). The next $2(d-2)+1$ are described by the state $\ket1$ and $2(d-2)$ states resulting from operations $R_y(\pi/2)$ and $R_x(3\pi/2)$ between levels $\ket1$ and $\ket j$ ($j=2,\dots,d-1$). To get the state $\ket1$, the transformation $R_x(\pi$) is applied to the initialization state $\ket0$. The similar procedure is performed for the states $\ket2,\dots,\ket{d-1}$. In total, $d^2$ different states are obtained in this way. These states form a basis for density matrices and are widely used in works on QPT \cite{mohseni2008,baldwin2014}.

At the output of a quantum process, each of these states is measured in the bases of a 2-level QST protocol. Each measurement circuit of the 2-level QPT protocol contains up to 3 elementary operations (2 for preparation and 1 for measurement) in each measurement circuit. The protocol is informationally complete.

\section{Accounting initialization and readout errors}\label{sect:init_readout}

In QPT, the input states and measurement operators in the case of ideal unitary operations are defined as follows:
\begin{equation}\label{eq:prep_meas}
  \rho_i = \mathcal{U}_i^p(\rho_0), \quad P_{ik} = \overline{\mathcal{U}}_i^m(\Pi_k),
\end{equation}
where $\mathcal{U}(\sigma)=U\sigma U^\dagger$, $\overline{\mathcal{U}}(\sigma)=U^\dagger\sigma U$. Operations $\mathcal{U}_i^p$ and $\mathcal{U}_i^m$ correspond to the preparation and basis change unitaries in $i$-th measurement circuit.

To account for initialization and readout errors in reconstruction model, one only needs to replace ideal $\rho_0$ and $\Pi_k$ with noisy ones in \eref{eq:prep_meas}. A good approximation is the diagonal form of these operators, which corresponds to the classical initialization and read errors:
\begin{equation}\eqalign{
  \rho_0 = \sum_{j=0}^{d-1}{a_j \ketbra{j}}, \quad \sum_{j=0}^{d-1}{a_j} = 1, \cr
  \Pi_k = \sum_{j=0}^{d-1}{b_{kj} \ketbra{j}}, \quad \sum_{k=0}^{d-1}{b_{kj}} = 1, \quad j = 0, \dots, d-1.
}
\end{equation}
This model is determined by a total of $d^2-1$ independent parameters. Since each complete readout of a qudit gives $d-1$ independent outcomes, the model requires implementing at least $d+1$ different circuits to determine all the parameters.

In the model we assume that the circuit contains initialization, unitary qudit transformation and readout. It turns out, however, that within the framework of this model it is impossible to design such a set of circuits that would allow for unambiguous estimation of all $a_j$ and $b_{kj}$. One can see this by first observing that in the case of all $a_j > 0$ one can always find $p$ such that $\rho_0 = \mathcal{E}_p(\rho_0^\prime)$. Since the depolarizing channel commutes with any unitary operation, the channel $\mathcal{E}_p$ can be classified as a readout error with operators $\Pi_k^\prime$. Thus, $\rho_0$ and $\Pi_k$ give the same set of probabilities as $\rho_0^\prime$ and $\Pi_k^\prime$. This makes these two sets indistinguishable in terms of measurement outcomes. This ambiguity is similar to the gauge invariance introduced in the gate set tomography \cite{matteo2020,blume_kohout2017,rudnicki2018}.

Due to this ambiguity, the number of circuits can be reduced from $d+1$ to $d$. To minimize errors, we will consider only circuits containing at most one elementary operation. The first circuit does not contain any operation and consists only of initialization and readout. The remaining $d-1$ circuits also contain an operation $R_x(\pi)$ between levels $\ket0$ and $\ket{j}$ ($j=1,\dots,d-1$) after initialization.

As an example, consider a qutrit ($d=3$) initialization error in the form of Gibbs distribution:
\begin{equation}\label{eq:init}
  a_j = e^{-\omega_j/T}/Z, \quad Z = \sum_j{e^{-\omega_j/T}}.
\end{equation}
Here $T = 1$ sets the effective system temperature, $\omega_0=0$, $\omega_1=4$, $\omega_2=6$ --- qutrit energy levels (in units of $T$). The readout error of the $j$-th level is simulated in the form
\begin{equation}\label{eq:readout}
  E_j = (1-b_0)\ketbra{j} + b_1(I-\ketbra{j}),
\end{equation}
where the coefficients $b_0 = 0.01$ and $b_1 = 0.02$ set the probabilities of false-negative and false-positive population readout result respectively. POVM operators are calculated from \eref{eq:readout} using \Eref{eq:readout_povm}. As before we assume that each 2-level operation is accompanied by depolarizing channel $\mathcal{E}_{0.001}$. The total sample size is 1,000,000.

The reconstruction of $a_j$ and $b_{kj}$ from the simulated measurements data was carried out by the maximum likelihood method. The optimization problem was solved using a genetic algorithm \cite{chernyavskiy2013} with open MATLAB library \cite{chernyavskiy_opt}. The following values were obtained:
\begin{equation}
  [\hat{a}_j] = \left(\begin{array}{c}0.97 \\ 0.03 \\ 0\end{array}\right), \quad
  [\hat{b}_{kj}] = \left(\begin{array}{ccc}0.98 & 0 & 0.01 \\ 0 & 1 & 0 \\ 0.02 & 0 & 0.99\end{array}\right).
\end{equation}

Based on these parameters, a tomography model is formed using \Eref{eq:prep_meas}. The resulting model is used to reconstruct the qudit unitary transformation accompanied by the channel $\mathcal{E}_{0.01}$. \Fref{fig:results}(b) shows the performance of the constructed model (``SPAM errors model 1''), the model based on true SPAM errors (``True model'') and the ideal SPAM model (``Ideal model''). The relatively low improvement of the new model can be explained by the ambiguity in determining $a_j$ and $b_{kj}$.

Let us consider a simpler parametrization corresponding to the true error model described in expressions \eref{eq:init} and \eref{eq:readout}. Such a model is specified by only three parameters: $T$, $b_0$, $b_1$ (the energy values of individual levels are assumed to be known). We consider a measurement set that does not include any operations: after initialization, the population of the $j$-th qudit level is read ($j=0,\dots,d-1$). The rest of the levels are not affected. Parameters estimations from the simulated experiment data (the total sample size is 1,000,000) are performed by the maximum likelihood method using a genetic optimization algorithm \cite{chernyavskiy2013,chernyavskiy_opt}. The resulting values were close to true ones: $\hat{T}=0.9868$, $\hat{b}_0=0.0116$, $\hat{b}_1=0.0203$. QPT results based on the model with these parameters (``SPAM errors model 2'') are shown in \Fref{fig:results}(b).

\section{Conclusion}

Systematic errors of quantum tomography significantly limit quantum state and quantum process reconstruction accuracy. Two fundamentally different approaches are error suppression and adequate accounting for existing errors. One of the ways to suppress errors is to use a minimum number of error prone operations in the tomography protocol. In this paper, we have presented a description of such a protocol for qudits of arbitrary dimension. In the case of a qubit, this protocol is equivalent to the 1-qubit MUB protocol.

However, even with a minimal number of operations, systematic errors can remain quite high. Usually, the most significant errors are due to error prone initialization and readout of the qudit state. Statistical estimation of these errors parameters allows one to construct a tomography model close to a real experiment. However, even in the case of classical form of initialization and readout errors, it turns out that it is impossible to get its unambiguous estimate. This ambiguity also results in quantum tomography systematic errors.

Thus, we have used the model of Gibbs distribution with non-linear parametrization and a small number of parameters. This approach made it possible to get an unambiguous parameters estimates from the simulated measurements data. We have shown that the resulting tomography model is capable of providing high fidelity.

\ack

This work was supported by the Program of activities of the leading research center ``Development of an experimental prototype of a hardware and software complex for the technology of quantum computing based on ions'' (Agreement No. 014/20) and by Theoretical Physics and Mathematics Advancement Foundation ``BASIS'' (Grant No. 20-1-1-34-1).

\section*{References}
\bibliography{article}

\end{document}